\documentstyle[12pt]{article}
\global\arraycolsep=2pt 
\input{epsf}
\begin{document}
\def\slash#1{#1 \hskip -0.5em / }
\thispagestyle{empty}
\begin{titlepage}

\begin{flushright}
CERN-TH.7162/94\\
hep-ph/9403249
\end{flushright}

\vspace{0.3cm}

\begin{center}
\Large\bf Higher Order $1/m$ Corrections    
          at Zero Recoil
\end{center}

\vspace{0.8cm}

\begin{center}
Thomas Mannel  \\
{\sl Theory Division, CERN, CH-1211 Geneva 23, Switzerland}
\end{center}

\vspace{0.8cm}

\begin{abstract}
\noindent
The general structure of the $1/m$ corrections at zero
recoil is studied. The relevant matrix elements 
are forward matrix elements of local higher dimensional operators
and their time ordered products with higher order terms from 
the Lagrangian. These matrix elements may be classified in a 
simple way and the analysis at the non recoil point for 
the form factor of heavy quark currents simplifies drastically. 
The second order recoil corrections to the 
form factor $h_{A1}$ of the axial vector current, 
relevant for the $|V_{cb}|$ determination from   
$B \to D^*$ decays, are estimated to be   
$-5\% < h_{A1} - 1 < 0$.
\end{abstract}
\vfill
\noindent
CERN-TH.7162/94\\
March 1994
\end{titlepage}

\setcounter{page}{1}
\section{Introduction}
Heavy quark effective theory is by now the standard description for 
systems with one heavy quark \cite{IW89}-\cite{MRR}.
The additional symmetries appearing in the limit of infinite heavy
quark mass yield model independent relations between form factors
appearing in the description of heavy hadron exclusive weak decays.
Aside from that,
heavy quark symmetries also yield statements about
the normalization of some form factors at zero recoil, i.e.\ the
point where the velocities of the initial and final hadrons are equal.
This fact has a very large phenomenological impact; it allows e.g.\
to perform a model independent determination of $|V_{cb}|$
by extrapolating to the endpoint of the lepton spectrum in the decay
$B \to D^* \ell \nu$.

QCD radiative corrections as well as recoil corrections have been
studied already to next-to-leading order \cite{corr}. While QCD radiative
corrections may be studied systematically, the recoil corrections in
general need new, non-perturbative input, which may be supplied, for
instance, by model estimates. For the case of the determination
of $|V_{cb}|$ from $B \to D^* \ell \nu$ the leading recoil corrections vanish 
at the non-recoil point due to
Lukes theorem \cite{Lu90} and the next-to-leading ones have been considered by
Falk and Neubert \cite{FN92}, who parameterized the form factors to order
$1/m_Q^2$, also off the point of equal velocities.

However, as will be discussed below, that the analysis
of the $1/m_Q$ corrections at zero recoil  simplifies
enormously, since then only forward matrix elements (i.e. matrix
elements of operators between mesons moving with the same velocity)
appear. In addition, the algebra of Dirac matrices simplifies and one
may obtain a simple expression for the next-to-leading
recoil corrections at the point of equal velocity. This expression
involves forward matrix elements of operators of higher dimension
and also time ordered products with higher order recoil terms from
the Lagrangian. The expressions we obtain have a simple interpretation, 
but its numerical evaluation needs input
beyond heavy quark effective theory.

Recently, the methods of the heavy mass expansion
have been applied also to inclusive decays by combining the
method of operator product expansion with heavy quark effective
theory \cite{CGG90}-\cite{Ma93}. 
This approach yields the heavy mass expansion
for decay rates and also for decay distributions; the leading term
in this expansion is the free quark decay rate and the corrections
may be studied systematically. Of course, the higher order corrections
need non-perturbative input, which is again forward matrix elements
of higher dimensional operators and time ordered products of
such operators with higher order terms from the Lagrangian. Thus 
the same matrix elements appear as in the model independent 
determination of $V_{cb}$.  

Finally, the relation of the heavy hadron mass to the mass of the
heavy quark is also given in terms of a $1/m_Q$ expansion. Higher
orders are again given by forward matrix elements of higher dimensional
operators needed as non-perturbative input to relate the heavy quark
mass with the heavy hadron mass.

In the present paper, a systematic study is performed for these
forward matrix elements appearing in all higher order calculations
at zero recoil, including the relevant
time ordered products with
higher order terms of the Lagrangian. It turns out that all the
forward matrix elements may be classified very simply and the
relevant matrix elements for calculations up to order $1/m_Q^3$
are given explicitly. 

The classification performed here allows to simplify the analysis 
of the recoil corrections to heavy quark weak decay form factors 
at $v=v'$ enormously, 
compared to the case off the equal velocity point. As an application 
the analysis for weak decay form factors is performed at the 
non-recoil point up to second 
order in the heavy mass expansion for the case of $b \to c$ transitions
and our results are compared with the ones obtained by Falk and Neubert
\cite{FN92}.

In the next section, a general discussion of the parametrization of 
the generic forward matrix element is given. It is split into three 
subsections. First we consider local higher dimensional operators 
in some detail and give numerical estimates for the matrix elements 
of operators up to dimension seven. In the second subsection we shall 
consider time ordered products with higher order terms from the 
Lagrangian. Finally, in the third subsection we consider the relation 
between the mass of the heavy meson and the mass of the heavy quark 
as a toy example, where the forward matrix elements play a role.

The general discussion of the forward matrix elements is then applied
in section 3 to the $1/m_Q^2$ corrections of the normalization
of the weak decay form factors in the decays $B \to D \ell \nu$ and
$B \to D^* \ell \nu$. The relevant form factors $h_+$ for 
$B \to D \ell \nu$ and $h_{A1}$ for $B \to D^* \ell \nu$ decays are 
discussed at the non-recoil point up to second order in the heavy mass 
expansion. We include also a semi-quantitative analysis ad estimate 
the size of the corrections relevant for the $V_{cb}$ determination 
to order $1/m_Q^2$.

\section{Higher Dimensional Operators and their Forward Matrix Elements}
Higher order terms in the heavy mass expansion of weak transition
matrix elements originate in general from two sources. The first
source is the heavy mass expansion of the operators for a heavy quark
$Q$ appearing in the weak transition Hamiltonian. At the matching scale
$m_Q$, this amounts to the replacement \cite{MRR}
\begin{equation} \label{>>>}
Q (x) = e^{-im_Q (vx)} \left( 1 + \left[ \sum_{k=0}^\infty
        \left(\frac{-ivD}{2m_Q}\right)^k \right]
              \frac{i\slash{D}^\perp }{2m_Q} \right) Q_v (x)
\end{equation}
where $v$ is the velocity of the heavy hadron and $Q_v$ is the
operator of a static heavy quark moving with velocity $v$.
Furthermore, it is convenient to define
\begin{equation} \label{perp}
D_\mu^\perp = (g_{\mu \nu} - v_\mu v_\nu  ) D^\nu
\quad vD^\perp = 0 .
\end{equation}
These terms in general lead to contributions, which are matrix elements
of local operators.

Secondly, also the Lagrangian of full QCD is expanded in $1/m_Q$
and the higher orders in $1/m_Q$ are treated as perturbations. This leads
to time ordered products involving these higher order terms of the
Lagrangian and the weak transition operator. The corrections
of higher orders
in the $1/m_Q$ expansion to the Lagrangian are given at tree
level by \cite{MRR}
\begin{equation} \label{lint}
{\cal L}_I = \sum_{j=1}^\infty {\cal L}_I^{(j)}
           = \sum_{j=1}^\infty \left(\frac{1}{2m_Q}\right)^j
             \bar{Q}_v (-i\slash{D}^\perp)
             (ivD)^{j-1} (i\slash{D}^\perp) Q_v
\end{equation}
In every order $j$ it is convenient to split ${\cal L}_I^{(j)}$ into 
a generalized kinetic energy operator ${\cal K}^{(j)}$ and  
chromomagnetic moment operator ${\cal G}^{(j)}$ defined as 
\begin{eqnarray}
{\cal K}^{(j)} &=& \bar{Q}_v (iD^\perp_\alpha)
             (-ivD)^{j-1} (iD^{\perp \alpha}) Q_v \\
{\cal G}^{(j)} &=& -i \bar{Q}_v (iD^\perp_\alpha)
             (-ivD)^{j-1} (iD^\perp_\beta) \sigma^{\alpha \beta} Q_v
\end{eqnarray}

The interpretation of these time ordered product terms is obvious.
The heavy hadron states of full QCD still depend on the heavy
quark mass, and this dependence is also treated in a $1/m_Q$
expansion. The leading term is the state taken in the infinite mass
limit, the ``static state'', which is the convenient one for
a heavy quark effective theory calculation, since it does not
depend on the heavy mass any more. The time ordered products 
account for correct mass dependence of the full QCD state, and the
matrix elements then have to be evaluated using the ``static'',
mass independent state.

\subsection{Local Operators of Higher Dimension}
The generic operator of dimension $n+3$
appearing in the contexts mentioned above
is of the form
\begin{equation} \label{genop}
{\cal O}^{(\Gamma)}_{\mu_1, \mu_2 \cdots \mu_n} =
\bar Q_v (iD_{\mu_1}) (iD_{\mu_2})
\cdots (iD_{\mu_n}) \Gamma Q_v
\end{equation}
where $\Gamma$ is an arbitrary Dirac matrix.

The Dirac matrix appearing in (\ref{genop}) may be expanded
into the 16 basis Dirac matrices $1$, $\gamma_5$, $\gamma_\mu$
$\gamma_5 \gamma_\mu$ and $\sigma_{\mu \nu}$. However, the
matrix $\Gamma$ is sandwiched between projectors
$$
P_+ = \frac{1}{2} (1+\slash{v})
$$
which are contained in the heavy quark fields $Q_v$. This
projection amounts to the replacements
\begin{eqnarray}
1 \longrightarrow P_+ = \frac{1}{2} (1+\slash{v})
&\qquad &
\gamma_\mu \longrightarrow P_+ \gamma_\mu P_+ = v_\mu P_+
\\
\gamma_\mu \gamma_5 \longrightarrow P_+ \gamma_\mu \gamma_5 P_+ = s_\mu
&\qquad &
\gamma_5 \longrightarrow P_+ \gamma_5 P_+ = 0
\\ \label{sigrep}
(-i) \sigma_{\mu \nu} \longrightarrow P_+ (-i) \sigma_{\mu \nu}P_+ =
     i v^\alpha \epsilon_{\alpha \mu \nu \beta} s^\beta
\end{eqnarray}
where we have defined the spin matrices $s_\mu$, which are the
generalizations of the Pauli matrices for the frame moving with
velocity $v$. They satisfy the relations
\begin{equation} \label{pauli1}
s_\mu s_\nu = (-g_{\mu \nu} + v_\mu v_\nu ) P_+
              + i \epsilon_{\alpha \mu \nu \beta} v^\alpha s^\beta
\qquad v \cdot s = 0
\end{equation}
Consequently, the Dirac matrix $\Gamma$ sandwiched between the
projectors may be expanded into the four matrices $1$ and
$s_\mu$
\begin{equation}
P_+ \Gamma P_+ = \frac{1}{2} P_+ \mbox{ Tr } \left\{ P_+ \Gamma \right\}
-\frac{1}{2} s_\mu \mbox{ Tr } \left\{ s^\mu \Gamma \right\} ,
\end{equation}
and it is sufficient to consider only the two operators
\begin{eqnarray}
{\cal O}^{(1)}_{\mu_1, \mu_2 \cdots \mu_n} &=&
\bar Q_v (iD_{\mu_1}) (iD_{\mu_2})
\cdots (iD_{\mu_n}) Q_v \\
{\cal O}^{(s)}_{\mu_1, \mu_2 \cdots \mu_n; \lambda} &=&
\bar Q_v (iD_{\mu_1}) (iD_{\mu_2})
\cdots (iD_{\mu_n}) s_\lambda Q_v
\end{eqnarray}

In the following we shall consider the matrix elements between
the ground state pseudoscalar and vector  mesons. There are two
different cases to be studied. In the first case the initial and
the final state are both either $0^-$ or $1^-$; in the second case
the initial state is $0^-$ and the final state is $1^-$ or vice
versa. All these different cases are related by heavy quark spin 
symmetry, which implies the relations 
\begin{equation}
Q_{v} | H(v) \rangle = \gamma_5 \slash{\epsilon} Q_{v}
       | H^*(v,\epsilon) \rangle \qquad
Q_{v} | H^*(v,\epsilon) \rangle = \gamma_5 \slash{\epsilon}
       Q_{v}| H(v) \rangle  .
\end{equation}
where $| H(v) \rangle $ and $| H^*(v,\epsilon) \rangle $ 
denotes the $0^-$ and $1^-$ ground state meson respectively. 

Further restrictions on the structure of the matrix elements
may be obtained from the equations of motion for the heavy quark 
and the gluons
\begin{equation}
ivD Q_v = 0 \qquad \left[ (iD^\mu) , [(iD_\mu), (iD_\nu)] \right]_{ab} =
4 \pi \alpha_s 
\sum_q \left[ \bar{q}_b \gamma_\mu q_a - \frac{1}{N_c} \delta_{ab}\, 
\bar{q}_d \gamma_\mu q_d \right]
\end{equation}
where the sum runs over all quark flavors and $a,b,d$ are color 
indices.

We shall start the discussion with the case where both, initial and
final state are either $0^-$ or $1^-$. Spin symmetry relates the 
matrix elements of $0^-$ mesons with the ones of the $1^-$ case 
in the following way
\begin{eqnarray} \label{SS}
\langle H(v) | {\cal O}^{(1)}_{\mu_1, \mu_2 \cdots \mu_n} 
| H(v) \rangle &=& \langle  H^*(v,\epsilon) |
{\cal O}^{(1)}_{\mu_1, \mu_2 \cdots \mu_n} | H^*(v,\epsilon) \rangle
\\ 
\langle H(v) | {\cal O}^{(s)}_{\mu_1, \mu_2 \cdots \mu_n; \lambda} 
| H(v) \rangle &=& - \frac{1}{3}
\langle  H^*(v,\epsilon) | 
{\cal O}^{(s)}_{\mu_1, \mu_2 \cdots \mu_n; \lambda}
| H^*(v,\epsilon) \rangle , 
\end{eqnarray}
and we shall consider in the following only the $0^-$ case.  

The equations of motions for the heavy quark imply that the matrix
elements of both ${\cal O}^{(1)}$ and ${\cal O}^{(s)}$ have to vanish, 
if the first or the last index, i.e. $\mu_1$ or $\mu_n$ is contracted 
with the velocity $v$. Furthermore, the spin vector $s$ is also 
orthogonal to the velocity and thus the matrix elements have to satisfy
\begin{eqnarray}
v^{\mu_1}\langle H(v) | {\cal O}^{(1)}_{\mu_1, \mu_2 \cdots \mu_n} 
| H(v) \rangle &=& 
v^{\mu_n} \langle H(v) | {\cal O}^{(1)}_{\mu_1, \mu_2 \cdots \mu_n} 
| H(v) \rangle = 0   \nonumber \\  
v^{\mu_1} \langle H(v) | 
{\cal O}^{(s)}_{\mu_1, \mu_2 \cdots \mu_n; \lambda} | H(v) \rangle 
&=& v^{\mu_n} \langle H(v) |
{\cal O}^{(s)}_{\mu_1, \mu_2 \cdots \mu_n; \lambda} | H(v) \rangle 
= 0 \nonumber \\ \label{restr} 
v^\lambda \langle H(v) | 
{\cal O}^{(s)}_{\mu_1, \mu_2 \cdots \mu_n; \lambda} | H(v) \rangle 
&=& 0 
\end{eqnarray}
Note that contractions with any other
index may be related to a gluon field strength
$[(iD_\mu), (iD_\nu)] = ig G_{\mu \nu}$, e.g.
\begin{equation}
(ivD) (iD_{\mu_n}) Q_v | H(v) \rangle =  
 -ig \, v^\alpha G_{\alpha \mu_n} Q_v
| H(v) \rangle ,
\end{equation}
and are thus in general nonzero.

Combining the information form the spin structure and the 
restrictions form the equation of motion 
of the heavy quark one obtains for the forward matrix 
element of ${\cal O}^{(1)}$ the 
general expression
\begin{equation}
\langle H(v) |
\bar Q_v (iD_\alpha)
(iD_{\nu_1}) \cdots (iD_{\nu_{n-2}}) (iD_\beta) Q_v
| H(v) \rangle  =  2 M_H [g_{\alpha \beta} - v_\alpha v_\beta ]
A_{\nu_1 \cdots \nu_{n-2}} 
\end{equation}
The tensor $A$ is constructed from
$g_{\mu_i \mu_j}$ and $v_{\mu_i}$. It is a simple combinatorical
exercise to show that the number $N$ of independent scalar parameters
is
\begin{equation}
N(n) = 1 + (n-2)! \sum_{k=1}^{[n]/2-1} \left(\frac{1}{2}\right)^k
    \frac{1}{(n-2(k+1))!}
\end{equation}
where $n>2$ and $[n] = n$ for $n$ even and $[n] = n-1$ for $n$ odd. 
The number of independent parameters grows rapidly, the first 
few are $N(4)=2$, $N(5)=4$, $N(6)=13$, $N(7)=41$ and $N(8)=196$. 
 
The matrix elements of ${\cal O}^{(s)}$ are parity odd quantities. 
The general form of these matrix elements, which is compatible with 
the restrictions (\ref{restr}), is given by 
\begin{eqnarray}
 \langle H(v) | {\cal O}_{\alpha \mu_1 \cdots \mu_{n-2} \beta; \lambda}
| H(v) \rangle
&=& 2M_H d_H i \varepsilon_{\nu \alpha \beta \lambda } v^\nu
    B_{\nu_1 \cdots \nu_{n-2}}  \\
&+& 2M_H d_H [g_{\alpha \beta} - v_\alpha v_\beta ] 
    C_{\nu_1 \cdots \nu_{n-2};\lambda}^{(1)}  \nonumber \\
&+& 2M_H d_H [g_{\alpha \lambda} - v_\alpha v_\lambda ] 
    C_{\nu_1 \cdots \nu_{n-2};\beta}^{(2)} \nonumber \\
&+& 2M_H d_H [g_{\beta \lambda} - v_\beta v_\lambda ] 
    C_{\nu_1 \cdots \nu_{n-2};\alpha}^{(3)} \nonumber 
\end{eqnarray}
where $d_H = 3$ for a pseudoscalar meson and $d_H = -1$ for 
a vector meson. The tensors $C^{(k)}$ are parity odd and 
vanish, if the last index in contracted with $v$. 
 
Up to dimension seven the number of parameters is still manageble, and 
some of them are more or less well known numerically. 
The only nonvanishing matrix element between heavy
meson states of the dimension 3 operators is
\begin{equation} \label{dim3}
\langle H(v) | \bar Q_v Q_v | H(v) \rangle
= 2M_H
\end{equation}
and its value is given by the choice of the normalization. 
Here $M_H$ is the mass of the heavy meson in the static limit. 

All matrix elements of the dimension 4 operators vanish due to
the equations of motion; all matrix elements of the dimension 5
operators are given in terms of two parameters $\lambda_1$
and $\lambda_2$
\begin{eqnarray}
\langle H(v) |
\bar Q_v (iD_\alpha) (iD_\beta) Q_v
| H(v) \rangle  &=& 2 M_H [g_{\alpha \beta} - v_\alpha v_\beta ]
\frac{1}{3}\lambda_1 \\
\langle H(v) |
\bar Q_v (iD_\alpha) (iD_\beta) s_\lambda Q_v
| H(v) \rangle
&=&   2 M_H d_H i \varepsilon_{\nu \alpha \beta \lambda } v^\nu
\frac{1}{6}\lambda_2
\end{eqnarray}
where the prefactors are chosen to comply with the definition in 
\cite{FN92}. The parameter $\lambda_2$ corresponds to the 
leading term in $1/m_Q$
for the mass splitting between the ground
state $1^-$ and $0^-$ mesons \cite{FGL92}, while the kinetic energy 
parameter $\lambda_1$ is not related 
in an easy way to a measurable quantity.
From QCD sum rule analyses one obtains values of 
$\lambda_1 = - 0.6 \pm 0.1$ GeV ${}^2$ \cite{Br92}, 
but these calculations have been 
criticized recently and a much lower value of $\lambda_1$
has been suggested using an improved sum rule technique \cite{Npriv}.
On the other hand,  
bounds have been derived in a quantum mechanical framework indicating
that $\lambda_1 < - 0.18$ GeV ${}^2$ \cite{Bmotion}. In the numerical 
studies presented below we shall vary $\lambda_1$ in some range 
and hence we shall use the values
\begin{equation} \label{lamnum}
-0.3 \mbox{ GeV}^2 < \lambda_1 < - 0.1 \mbox{ GeV}^2 \qquad 
\lambda_2 =  0.12 \mbox{ GeV}^2 .
\end{equation}
The parameter $\lambda_2$ is scale dependent and we
define $ \lambda_2 = \lambda_2 (m_b)$.

The matrix elements of the dimension 6 operators are also given
in terms of only two parameters $\rho_1$ and $\rho_2$
\begin{eqnarray}
\langle H(v) |
\bar{Q}_v (iD_\alpha) (iD_\mu)  (iD_\beta) h_v
| H(v) \rangle   
&=&  2 M_H [g_{\alpha \beta} - v_\alpha v_\beta ] v_\mu
\frac{1}{3}\rho_1 \\
\langle H(v) |
\bar{Q}_v (iD_\alpha) (iD_\mu) (iD_\beta) s_\lambda Q_v
| H(v) \rangle 
&=& 2M_H d_H i \varepsilon_{\nu \alpha \beta \lambda } v^\nu v_\mu
\frac{1}{6} \rho_2 .
\end{eqnarray}
In order to estimate $\rho_1$ we may employ the equations of motion
for the gluon fields and relate this parameter to a forward matrix 
element of a four fermion operator
\begin{equation} \label{fourferm}
-4 M_H \rho_1  =  4 \pi \alpha_s \sum_q 
\langle H(v) | \left[ (\bar{Q}_{v,a} Q_{v,b})
(\bar{q}_b \slash{v} q_a ) - 
\frac{1}{N_c}  (\bar{Q}_{v,a} Q_{v,b})
  (\bar{q}_b \slash{v} q_b ) \right]
| H(v) \rangle 
\end{equation}
where $a$ and $b$ are color indices. Contracting 
(\ref{fourferm}) with $v$ we may rewrite the matrix element as an
interaction between two vector currents. Using the Fierz theorem 
we rearrange the quark fields in order to apply vaccum 
insertion, after which one is left with matrix elements of heavy 
light operators between the heavy meson and vacuum. These matrix 
elements are all related to the heavy meson decay constant $f_H$ 
due to  heavy quark spin symmetry. The 
estimate for the parameter $\rho_1$ reads under these assumptions
\begin{equation} \label{rhores}
\rho_1 = \frac{1}{2} \pi \alpha_s  \frac{N_c^2 - 1}{N_c^2}  f_H^2 M_H .
\end{equation}
A similar estimate has been performed in \cite{Bmotion}.

However, (\ref{rhores}) has the usual problem of a matrix element 
after factorization. The 
original matrix element defining $\rho_1$ is expected to have a 
different behavior under renormalization group transformations 
as the result after factorization. In other words, one has to 
define at which scale factorization is performed. We shall factorize
the matrix elements at the scale $m_b$ and thus use the following 
set of parameters:  $\alpha_s = 
\alpha_s (m_b) = 0.2$ and $M_H = 5.28$ GeV. Varying the heavy 
meson decay constant between 150 and 200 MeV we obtain 
\begin{equation} 
( \rho_1 )^{1/3}  = (300 - 450) \mbox{ MeV}.
\end{equation}
This number is of the same size as e.g. $\lambda_2^{1/2} = 350$ MeV.

Finally, the forward matrix elements of the dimension 7 operators 
${\cal O}^{(1)}$ may be written in terms
of two parameters $\eta$ and $\tau$
\begin{equation}
\langle H(v) | {\cal O}^{(1)}_{\alpha \mu_1 \mu_2 \beta}
| H(v) \rangle  
 =  2 M_H \frac{1}{3}[g_{\alpha \beta} - v_\alpha v_\beta ]
(g_{\mu_1 \mu_2} \eta_1 - v_{\mu_1} v_{\mu_2} \tau_1 )  
\end{equation}
while the general form of the dimension seven operator 
${\cal O}^{(s)}$ is more complicated 
\begin{eqnarray}
\langle H(v) | {\cal O}^{(s)}_{\alpha \mu_1 \mu_2 \beta ; \lambda}
| H(v) \rangle 
&=& - 2M_H d_H i 
\varepsilon_{\alpha \beta \lambda \nu} v^\nu 
(g_{\mu_1 \mu_2} B_1 - v_{\mu_1} v_{\mu_2} B_2 )
\\ 
&& +  2M_H d_H C^{(1)} 
                 [g_{\alpha \beta} - v_\alpha v_\beta ] 
\varepsilon_{\rho \mu_1 \mu_2 \lambda} 
\nonumber \\
&&  +  2M_H d_H C^{(2)} 
                 [g_{\alpha \lambda} - v_\alpha v_\lambda ] 
\varepsilon_{\rho \mu_1 \mu_2 \beta }
\nonumber \\
&&  +  2M_H d_H C^{(3)} 
                 [g_{\lambda \beta} - v_\lambda v_\beta ] 
\varepsilon_{\rho \mu_1 \mu_2 \alpha} 
\nonumber  .
\end{eqnarray}

One may again apply the equations of motion for the gluon field
to relate this to a matrix element involving the light quark current.
In this way one obtains a relation of the form 
\begin{eqnarray}
&& 2 M_H  (4 \eta + \tau) (g_{\alpha \beta } -  v_\alpha v_\beta ) 
\\ \nonumber 
&& = - 4 \pi \alpha_s \sum_q 
\langle H(v) | \left[ ((i D_\alpha  \bar{Q}_{v,a}) Q_{v,b})
(\bar{q}_b \gamma_\beta q_a) - 
\frac{1}{N_c} ((i D_\alpha  \bar{Q}_{v,a}) Q_{v,a})
(\bar{q}_b \gamma_\nu q_b ) \right] 
| H(v) \rangle 
\end{eqnarray}
This may again be estimated by using the Fierz theorem and vacuum
insertion. After factorization, using  
\begin{equation}
\langle H(v) | (i D_\alpha  \bar{Q}_v) \gamma_\mu \gamma_5 q | 0 \rangle 
= 3\bar{\Lambda} f_H M_H [ g_{\alpha \mu} - v_\alpha v_\mu ]
\qquad 
\bar{\Lambda} = M_H - m_Q , 
\end{equation}
one obtains 
\begin{equation}
4 \eta + \tau = 
6 \pi \alpha_s \frac{N_c^2 - 1}{N_c^2} \bar{\Lambda} f_H^2 M_H
\end{equation}
 
With the same of parameters and under the same assumptions 
as above one obtains the estimate
\begin{equation}
(4 \eta + \tau)^{1/4} =  (700 - 950) \mbox{ MeV},
\end{equation}
where $\bar{\Lambda}$ has been varied between 400 and 600 MeV. 
 
From these estimates it seems that the heavy mass expansion works 
quite well, at least at the non-recoil point. All the parameters 
up to dimension seven behave like the appropriate power of some 
small scale $\Lambda \sim$ 200 -- 500 MeV which means that the 
expansion in powers of $\Lambda / m_Q$ indeed has coefficients of 
order unity.   

The second case to be studied are matrix elements with a $0^-$
meson in the initial state and a  $1^-$ in the final state, or 
vice versa. These matrix elements do not introduce any new 
parameters, since they are related to the ones considered above
by heavy quark spin symmetry. However, one has to be a little 
more careful in this case, because one has to rotate the spin 
of only one of the heavy quarks. The forward matrix elements, 
which are considered here, involve only one velocity sector 
of heavy quark effective theory, and spin symmetry is a symmetry 
holding separately 
in each velocity sector. In order to rotate only the initial or 
the final state heavy quark spin, one has to 
choose in a first step two different velocities for initial 
and final state,  perform the spin rotation of one of the states 
using (\ref{SS}) in the corresponding velocity sector, and 
afterwards take the limit $v' \to v$. In this way one obtains 
the relations
\begin{eqnarray}
&& \langle H(v) |
\bar Q_v
(iD_{\mu_1}) \cdots (iD_{\mu_n})  Q_v
| H(v) \rangle =  -
\langle H(v) |
\bar Q_v
(iD_{\mu_1}) \cdots (iD_{\mu_n})  (s \epsilon) Q_v
| H^*(v,\epsilon) \rangle \qquad
\\ 
&& \langle H(v) |
\bar Q_v
(iD_{\mu_1}) \cdots (iD_{\mu_n})  Q_v
| H^*(v,\epsilon) \rangle =  -
\langle H(v) |
\bar Q_v
(iD_{\mu_1}) \cdots (iD_{\mu_n})  (s \epsilon) Q_v
| H (v) \rangle , \qquad 
\end{eqnarray}
relating the matrix elements of ${\cal O}^{(1)}$ between two $0^-$
or two $1^-$ states to the ones of 
${\cal O}^{(s)}$ between $0^-$ and a $1^-$ state, and vice versa.

\subsection{The Time Ordered Products with the Lagrangian}

The second type of matrix elements appearing in an analysis of 
higher order $1/m_Q$ corrections at zero recoil are time ordered 
products of the local operators discussed above and the terms 
appearing in the heavy mass expansion of the Lagrangian. 

We shall first consider the case of two different flavors $q_v$ 
and $Q_v$. The simplest terms are the two point matrix elements 
\begin{eqnarray}
&& (-i) \int d^4 x \, 
\langle H_q (v) | T \left[ \bar q_v
(iD_{\mu_1}) \cdots (iD_{\mu_n})  Q_v \, {\cal K}_Q^{(j)} (x) \right]
| H_Q (v) \rangle  \\
&& (-i) \int d^4 x \, 
\langle H_q (v) | T \left[ \bar q_v
(iD_{\mu_1}) \cdots (iD_{\mu_n})  Q_v \, {\cal G}_Q^{(j)} (x) \right]
| H_Q (v) \rangle 
\end{eqnarray}
where the operators without argument have to be taken at $x=0$.
${\cal K}_Q$ and ${\cal G}_Q$ are the kinetic and chromomagnetic 
terms for the quark $Q$ as defined above. 

The spin structure of the simplest two point matrix elements may 
be analyzed in the trace formalism 
\begin{eqnarray} \label{gmm}
&& (-i) \int d^4 x \, 
\langle H_q(v) | T \left[ \bar q_v \Gamma  Q_v \, 
               {\cal K}_Q^{(j)} (x) \right]
| H_Q (v) \rangle = - A  \,\,
\mbox{Tr } \left\{ \bar{M}(v) \Gamma M(v) \right\} \\
&& (-i) \int d^4 x \, 
\langle H_q (v) | T \left[ \bar q_v \Gamma  Q_v \, 
               {\cal G}_Q^{(j)} (x) \right]
| H_Q (v) \rangle =  \frac{1}{2} B \,\, \nonumber 
\mbox{Tr } \left\{ (-i )\sigma_{\alpha \beta}
\bar{M}(v) P_+ (-i) \sigma^{\alpha \beta} M (v) \right\} 
\end{eqnarray}
where $\Gamma$ is a general Dirac matrix, which is a linear combination 
of $1$ and $s_\mu$, and $M(v)$ are the usual representation matrices for 
the heavy ground state mesons  
\begin{equation}
M(v) = \frac{1}{2} \sqrt{M_H} \left\{ \begin{array}{ll}
(\slash{v} +1) \gamma_5 &
\mbox{pseudoscalar meson} \\
- (\slash{v} +1) \slash{\epsilon} &
\mbox{vector meson, polarization }\epsilon
\end{array} \right. , 
\end{equation}
and the normalization is chosen according to (\ref{dim3}).

The matrix $\sigma_{\alpha \beta}$ in the expression for the chromomagnetic 
moment operator appears only between projection operators $P_+$ and it
is convenient to switch to a representation using the Pauli matrices
(\ref{sigrep}). In this representation one has 
\begin{equation}
{\cal G}^{(j)} = i v_\mu \epsilon^{\mu \alpha \beta \lambda }
                 \bar{Q}_v (iD_\alpha)
             (ivD)^j (iD_\beta) s_\lambda  Q_v
\end{equation}
and we write for the second equation of (\ref{gmm})
$$
(-i) \int d^4 x \, 
\langle H_q (v) | T \left[ \bar q_v \Gamma  Q_v \, 
               {\cal G}_Q^{(j)} (x) \right]
| H_Q (v) \rangle =  - B \,\,
\mbox{Tr } \left\{ \gamma_\lambda \gamma_5  
\bar{M(v)} \Gamma s^\lambda M(v) \right\} 
$$
 
The representation in terms of the Pauli matrices
is very useful, as soon as more than one insertion 
of a chromomagnetic moment operator appears, since the spin structure
of products of chromomagnetic operators correspond to products of 
the spin matrices $s$ which may be reduced using the relation 
(\ref{pauli1}). For example, the product of two chromomagnetic 
moment operator insertion may be written as
\begin{eqnarray}
&& (-i)^2 \int d^4 x \, d^4 y \, \nonumber
\langle H_q(v) | T \left[ \bar q_v Q_v \, {\cal G}_Q^{(j)} (x)
          {\cal G}^{(j)}_Q (y) \right]
| H_Q (v) \rangle =    
 - \mbox{Tr } 
\left\{ {\cal T}_{\alpha \mu} 
\bar{M(v)} s^\alpha s^\mu M(v) \right\} 
\\
&& =  - \mbox{Tr } 
\left\{ {\cal T}^{\alpha \mu} (-g_{\alpha \mu} + v_\alpha v_\mu )  
\bar{M(v)} M(v) \right\} 
- \mbox{Tr } 
\left\{ {\cal T}^{\alpha \mu} i\epsilon_{\rho \alpha \mu \nu} v^\rho
\bar{{\cal H}(v)} s^\nu {\cal H}(v) \right\} 
\end{eqnarray} 
where ${\cal T}$ parameterizes the light degrees of freedom 
\begin{equation}
{\cal T}_{\alpha \beta} = \frac{1}{3} T^{(1)} 
                   (v_\alpha v_\beta - g_{\alpha \beta}) + 
                   \frac{i}{2} T^{(2)} 
                   \epsilon_{\mu \alpha \beta \lambda} 
                   v^\mu \gamma^\lambda \gamma_5  ,
\end{equation}
and one obtains 
$$  
(-i)^2 \int d^4 x \, d^4 y \, 
\langle H_q (v) | T \left[ \bar q_v Q_v \, {\cal G}^{(j)} (x)
          {\cal G}_Q^{(j)} (y) \right]
| H_Q (v) \rangle = 2M_H (T^{(1)} + d_H T^{(2)})
$$
In this way 
one may easily identify the spin symmetry conserving and 
spin symmetry violating contributions of such products. 

The equations of motion also imply restrictions on the 
matrix elements of time ordered products \cite{WM93,FLS93}. In principle,
one obtains the same relations as for the local terms, for example 
$$ 
\langle H_q (v) | T \left[ \bar q_v (ivD)  Q_v \, {\cal K}_Q^{(j)} (x) \right]
| H_Q (v) \rangle = 0  
$$
However, there may be an ambiguity depending whether the derivative 
acts on the $T$ symbol or not. If one also takes the derivative of the 
step functions comming from the $T$ symbol, then one obtains a local 
contribution of the form 
$$ 
\langle H_q(v) | T \left[ \bar q_v (iD_\mu)  Q_v \, {\cal K}^{(j)} (x) \right]
| H_Q(v) \rangle \sim i \delta^4 (x) v_\mu  
\langle H_q(v) | \bar q_v   
(iD^\perp_\alpha) (ivD)^j (iD^{\perp \alpha}) Q_v
| H_Q(v) \rangle .     
$$
which may in general be reabsorbed into a redefinition of the $T$ product.
However, in the applications discussed 
below only the perpendicular components of the derivatives defined in 
(\ref{perp}) enter the expressions as e.g.
$$ 
\langle H_q(v) | T \left[ \bar q_v (i\slash{D}^\perp)  Q_v \, 
               {\cal K}_Q^{(j)} (x) \right]
| H_Q(v) \rangle = 0   , 
$$
and hence there will be no contribution 
from such terms. 

Finally, the flavor diagonal case may be discussed by inserting first 
the correction terms for the Lagrangian of the quark $q$ 
$$
{\cal L}^{(j)}= \left(\frac{1}{m_Q}\right)^j 
                \left( {\cal K}_Q^{(j)}+ {\cal G}_Q^{(j)} \right) 
               + 
                \left(\frac{1}{m_q}\right)^j 
                \left( {\cal K}_q^{(j)}+ {\cal G}_q^{(j)} \right) 
$$
and then consider the case $q=Q$. In this case one has insertions 
in both lines, the one corresponding to $q$ and to $Q$. When the 
masses are equal, both insertions are parametrized by the same form 
factor. However, the spin structure is different; in particular, the 
insertion of the chromomagnetic moment operator yields a Pauli 
matrix $s$ to the right of $\Gamma$ for $Q$ , while $s$ occurs 
to the left of $\Gamma$ for $q$. Thus one obtains for the examples 
studied above
\begin{eqnarray} \label{gmmfd}
&& (-i) \int d^4 x \, 
\langle H_Q(v) | T \left[ \bar Q_v \Gamma  Q_v \, 
               {\cal K}^{(j)} (x) \right]
| H_Q (v) \rangle = - 2 A  \,
\mbox{Tr } \left\{ \bar{M}(v) \Gamma M(v) \right\} \\ 
&& (-i) \int d^4 x \, 
\langle H_Q (v) | T \left[ \bar Q_v \Gamma  Q_v \, 
               {\cal G}^{(j)} (x) \right]
| H_Q (v) \rangle =  - B \,\,
\mbox{Tr } \left\{ \gamma_\lambda \gamma_5  
\bar{M(v)} \{ \Gamma \, , \, s^\lambda \} M(v) \right\} 
\end{eqnarray}
where $\{ , \}$ denotes the anticommutator of the two Dirac matrices.

\subsection{Simple Application: The Heavy Meson Mass}

The mass of a heavy hadron may be expanded in inverse powers of the 
heavy quark mass. The lowest order terms of this expansion have been 
considered and one may extend this analysis to higher orders
using the above discussion of the forward matrix elements.  

The relation between the heavy meson mass $m_H$ and the mass of the heavy 
quark is given by
\begin{equation}
m_H = M_H - \frac{1}{2M_H}
\langle H(v)|  T \left[ {\cal L}_I (0) \exp \left(
-i \int d^4 x \, {\cal L}_I (x) \right) \right]
| H(v) \rangle 
\end{equation}
where $M_H = m_Q + \bar{\Lambda}$ is the mass of the heavy hadron 
in the limit $m_Q \to \infty$ 

The $1/m_Q$ expansion of the hadron mass is obtained by 
inserting the expression
(\ref{lint}) into the time ordered product. Up to order $1/m_Q^2$
one finds
\begin{eqnarray}
&& m_H = m_Q + \bar{\Lambda} - \frac{1}{2 M_H} \sum_{j=1,2} 
\left(\frac{1}{2m_Q}\right)^j \left[ 
\langle H(v)| {\cal K}^{(j)} (0) | H(v) \rangle +
\langle H(v)| {\cal G}^{(j)} (0) | H(v) \rangle \right]
\nonumber \\ \nonumber
&&- (-i) \frac{1}{2M_H} \left( \frac{1}{2m_Q} \right)^2 
\int d^4 x \, \langle H(v)|  
T \left[ \left( {\cal K}^{(1)} (0) + {\cal G}^{(1)} (0) \right) 
         \left( {\cal K}^{(1)} (x) + {\cal G}^{(1)} (x) \right)
\right] | H(v) \rangle 
\\ 
&&+ {\cal O} (1/m_Q^3)
\end{eqnarray}
The matrix elements appearing here are exactly of the type
considered above. To order $1/m_Q$ there are the two parameters
$\lambda_1$ and $\lambda_2$, while to order $1/m_Q^2$ one has 
not only local operators, but also time ordered products to consider.
The two local matrix elements are given in terms of $\rho_1$ and 
$\rho_2$, while the time ordered products are parameterized according 
to 
\begin{eqnarray}
(-i) \int d^4 x \, \langle H(v)|  
T \left[  {\cal K}^{(1)} (0) {\cal K}^{(1)} (x) 
\right] | H(v) \rangle &=& - 2 T_1 \mbox{Tr } \left\{ 
           \bar{M}(v)  M(v) \right\}
\\
(-i) \int d^4 x \, \langle H(v)|  
T \left[  {\cal K}^{(1)} (0) {\cal G}^{(1)} (x) 
\right] | H(v) \rangle &=& - 2 T_2 \mbox{Tr } 
           \left\{ \gamma_\lambda \gamma_5 
           \bar{M} (v) s^\lambda M(v) \right\}
\\  
(-i) \int d^4 x \, \langle H(v)|  
T \left[  {\cal G}^{(0)} (0) {\cal G}^{(0)} (x) 
\right] | H(v) \rangle &=& - \mbox{Tr } 
       \left\{ T^{\alpha \beta}  
       \bar{M} (v) \{ s_\alpha \, , \, s_\beta \} M(v)  \right\}
\end{eqnarray}
where 
\begin{equation}
T_{\alpha \beta} = \frac{1}{3} T_3^{(1)} 
                   (v_\alpha v_\beta - g_{\alpha \beta}) + 
                   \frac{i}{2} T_3^{(2)} 
                   \epsilon_{\mu \alpha \beta \lambda} 
                   v^\mu \gamma^\lambda \gamma_5 
\end{equation} 
Using this parametrization one obtains
\begin{equation}
m_H = M_H - \frac{1}{2m_Q} (\lambda_1 + d_H \lambda_2 ) 
- \left(\frac{1}{2m_Q}\right)^2 [\rho_1 + 2 T_1 + 2 T_3^{(1)} 
                           + d_H (\rho_2 + 2 T_2 ) ] 
+  {\cal O} (1/m_Q^3)
\end{equation}
where the spins symmetry breaking contribution of the double 
insertion  of the chromomagnetic moment operator does not 
contribute, since we are dealing with the 
flavor diagonal case. 

The parameter $\rho_1$ has been estimated above; this term 
contribute only about 0.5 MeV to the mass of the $B$ meson. The 
parameter $\rho_2$ is more difficult to estimate, but a reasonable 
guess is certainly $|\rho_2| \sim |\rho_1|$, motivated by the fact that 
$|\lambda_1| \sim |\lambda_2|$. 

The time ordered products are much harder to estimate. They require 
in general a model describing the dynamics of the light degrees of 
freedom. We shall not consider this here, but it seems reasonable 
that the time ordered products are of similar magnitude than as the 
local terms.

\section{Higher Order Corrections to the $V_{cb}$
         Determination}
As the main application we consider the higher order
corrections to the semileptonic decay of a $B$ meson into a $D$
or $D^*$ meson. In this case we have to deal with to heavy flavors
$b$ and $c$, and the corresponding static operators are denoted $b_v$
and $c_v$ respectively. These higher order terms have been considered
already in \cite{FN92} also off the non-recoil point; however at the
point $v=v'$ the analysis simplifies drastically compared to the one 
off the non-recoil point.

The form factors to be considered are the ones of the vector and the 
axial vector current, defined by
\begin{eqnarray} \label{vector}
\langle B(p) | \bar{b}\gamma_\mu  c | D (p') \rangle  
&=& \sqrt{m_B m_{D} }\, h_+ (vv') 2 v_\mu + \cdots 
\\ \label{axial}
\langle B(p) | \bar{b}\gamma_\mu \gamma_5 c | D^* (p', \epsilon) \rangle  
&=& \sqrt{m_B m_{D^*} }\, h_{A1} (vv') (1+ vv') \epsilon_\mu + \cdots 
\end{eqnarray}
where the ellipses denote terms which vanish as $v \to v'$ due to 
their kinematic prefactors. 
Here $b$ and $c$ are the fields of full QCD and $| B(p) \rangle $ and $|D(p')\rangle $ are the full QCD states. Both 
form factors $h_+$ and $h_{A1}$ are normalized at the non-recoil point
$v=v'$ in the heavy quark limit such that $h_+ = h_{A1} = 1$. In 
addition to these we also consider the matrix element 
\begin{equation} 
\langle B^*(p,\epsilon) | \bar{b}\gamma_\mu  c | D^* (p', \epsilon') \rangle  
= \sqrt{m_{B^*} m_{D^*} } h_1 (vv')  (-\epsilon \epsilon ') v_\mu + \cdots 
\end{equation}
which we shall need to derive normalization conditions. 

Using the $1/m_Q$ expansion (\ref{>>>}) for both operators $b$ and $c$
the contributions to the matrix element at the non-recoil point may be
classified into three species
\begin{equation}
\langle H_b (p) | \bar{b} \Gamma c | H_c (p') \rangle |_{v=v'} = L + T + M
+ {\cal O} (1/m_c^3)
\end{equation}
where $\Gamma = \gamma_\mu, \gamma_\mu \gamma_5$ and $H_b$ and $H_c$
are $B$, $B^*$ or $D$, $D^*$ respectively. 

The contribution 
$L$ are all local terms,
i.e. the ones which do not contain any time ordered product. They
originate from the expansion of the operators (\ref{>>>}) and read
\begin{eqnarray}
L &=& \langle H_b (v) | \bar{b}_v \Gamma c_v | H_c (v) \rangle
\\
&+& \left(\frac{1}{2m_c} \right)
\langle H_b (v) | \bar{b}_v \Gamma (i\slash{D}^\perp) c_v | H_c (v) \rangle
- \left(\frac{1}{2m_b} \right)
\langle H_b (v) | \bar{b}_v
\stackrel{\longleftarrow}{(i\slash{D}^\perp)}
\Gamma c_v | H_c (v) \rangle
\nonumber \\
&-& \left(\frac{1}{2m_c} \right)^2
\langle H_b (v) | \bar{b}_v \Gamma (ivD)
              (i\slash{D}^\perp) c_v | D(v) \rangle
- \left(\frac{1}{2m_b} \right)^2
\langle H_b (v) | \bar{b}_v
\stackrel{\longleftarrow}{(i\slash{D}^\perp)}
\stackrel{\longleftarrow}{(ivD)}
\Gamma c_v | H_c (v) \rangle 
\nonumber \\
&-& \left(\frac{1}{4m_b m_c} \right)
\langle H_b (v) | \bar{b}_v
\stackrel{\longleftarrow}{(i\slash{D}^\perp)}
\Gamma (i\slash{D}^\perp)  c_v | H_c (v) \rangle
+ {\cal O}(1/m^3)
\nonumber
\end{eqnarray}
where now $| H_b (v) \rangle$ and $| H_c (v) \rangle $ are the states in the
infinite mass limit.

From the discussion of the forward matrix elements it follows
that only the last term does not vanish. The terms of first order
in $1/m_Q$ are forward matrix elements of a dimension four operator
and hence zero, the terms of order
$1/m_b^2$ and $1/m_c^2$ vanish after a partial integration, which
for the forward matrix elements does not yield a surface term.
Only the mixed term of order $1/(m_b m_c)$ yields a contribution,
which may be related to $\lambda_1$ and $\lambda_2$.

The second class of terms are the time ordered products of the
current to leading order with the terms of order $1/m$ and $1/m^2$
of the Lagrangian. One obtains
\begin{eqnarray}
T &=& (-i) \left( \frac{1}{2m_c} \right) \int d^4 x \,
\langle H_b (v) | T \left[ \bar{b}_v \Gamma c_v {\cal L}_c^{(1)} (x) \right]
                   | H_c (v) \rangle
\\
 &+& (-i) \left( \frac{1}{2m_b} \right) \int d^4 x \,
\langle H_b (v) | T \left[ {\cal L}_b^{(1)} (x) \bar{b}_v \Gamma c_v  \right]
                   | H_c (v) \rangle
\nonumber \\
&+& (-i) \left( \frac{1}{2m_c} \right)^2 \int d^4 x \,
\langle H_b (v) | T \left[ \bar{b}_v \Gamma c_v {\cal L}_c^{(2)} (x) \right]
                   | H_c (v) \rangle
\nonumber \\
 &+& (-i) \left( \frac{1}{2m_b} \right)^2 \int d^4 x \,
\langle H_b (v) | T \left[ {\cal L}_b^{(2)} (x) \bar{b}_v \Gamma c_v  \right]
                   | H_c (v) \rangle
\nonumber \\
&+& \frac{(-i)^2}{2} \left( \frac{1}{2m_c} \right)^2 \int d^4 x \,d^4y \,
\langle H_b (v) | T \left[ \bar{b}_v \Gamma c_v {\cal L}_c^{(1)} (x)
                                      {\cal L}_c^{(1)} (y) \right]
                   | H_c (v) \rangle
\nonumber \\
&+& \frac{(-i)^2}{2} \left( \frac{1}{2m_b} \right)^2 \int d^4 x \,d^4y \,
\langle H_b (v) | T \left[ {\cal L}_b^{(1)} (x)
                        {\cal L}_b^{(1)} (y) \bar{b}_v \Gamma c_v \right]
                   | H_c (v) \rangle
\nonumber \\
&+& (-i)^2 \left( \frac{1}{4m_b m_c} \right)
\int d^4 x \,d^4y \,
\langle H_b (v) | T \left[  {\cal L}_b^{(1)} (x) \bar{b}_v \Gamma c_v
                                      {\cal L}_c^{(1)} (y) \right]
                   | H_c (v) \rangle
\nonumber
\end{eqnarray}
where 
here and in the following we suppress the argument of the current 
$\bar{b}_v \Gamma c_v$ which is $x=0$.
 
Finally, there are the mixed contributions $M$ containing a first
order term of the expansion of the operators (\ref{>>>})  and 
a first order term
of the Lagrangian
\begin{eqnarray}
M &=& (-i) \left( \frac{1}{2m_c} \right)^2 \int d^4 x \,
\langle H_b (v) | T \left[ \bar{b}_v \Gamma (i\slash{D}^\perp )
c_v {\cal L}_c^{(1)} (x) \right]
                   | H_c (v) \rangle
\\
 &-& (-i) \left( \frac{1}{2m_b} \right)^2 \int d^4 x \,
\langle H_b (v) | T \left[ {\cal L}_b^{(1)} (x) \bar{b}_v
\stackrel{\longleftarrow}{(i\slash{D}^\perp)} \Gamma
                          c_v \right]  | H_c (v) \rangle
\nonumber \\
&+& (-i) \left( \frac{1}{4m_c m_b} \right) \int d^4 x \,
\langle H_b (v) | T \left[ {\cal L}_b^{(1)} (x) 
\bar{b}_v \Gamma (i\slash{D}^\perp) c_v  \right]
                   | H_c (v) \rangle
\nonumber \\
&-& (-i) \left( \frac{1}{4m_b m_c} \right) \int d^4 x \,
\langle H_b (v) | T \left[ \bar{b}_v
\stackrel{\longleftarrow}{(i\slash{D}^\perp)} \Gamma
c_v {\cal L}_c^{(1)} (x) \right]
                   | H_c (v) \rangle
\nonumber
\end{eqnarray}
As discussed in section 2.2, these mixed terms all vanish due to the equations
of motion. 

In order to proceed further with the time ordered products one
has to split the Lagrangians ${\cal L}^{(i)}_{b/c}$ into its kinetic
and magnetic terms. In this way one may analyze the spin structure
of terms involving products of chromomagnetic moment operators by
employing
the trace formalism, and by using the fact that all products of Dirac
matrices may be reduced using the algebra of the Pauli matrices,
eq.(\ref{pauli1}).
The trace formalism gives for the terms of order $1/m$
\begin{eqnarray}
(-i) \int d^4 x \,
\langle H_b (v) | T \left[ \bar{b}_v \Gamma c_v {\cal K}_c^{(1)} (x) \right]
                   | H_c (v) \rangle &=&
- \chi_1 \mbox{ Tr } \left\{\bar{M} (v) \Gamma M(v) \right\}
\\ \nonumber 
(-i) \int d^4 x \,
\langle H_b (v) | T \left[ \bar{b}_v \Gamma c_v {\cal G}_c^{(1)} (x) \right]
                   | H_c (v) \rangle &=&
- \chi_3 \mbox{ Tr } \left\{\gamma_\lambda \gamma_5
                     \bar{M} (v) \Gamma s^\lambda M(v) \right\}
\\ \nonumber
(-i)  \int d^4 x \,
\langle H_b (v) | T \left[ \bar{b}_v \Gamma c_v {\cal K}_b^{(1)} (x) \right]
                   | H_c (v) \rangle &=&
- \chi_1 \mbox{ Tr } \left\{\bar{M} (v) \Gamma M(v) \right\}
\\ \nonumber
(-i)  \int d^4 x \,
\langle H_b (v) | T \left[ \bar{b}_v \Gamma c_v {\cal G}_b^{(1)} (x) \right]
                   | H_c (v) \rangle &=&
- \chi_3 \mbox{ Tr } \left\{\gamma_\lambda \gamma_5
                     \bar{M} (v) s^\lambda \Gamma M(v) \right\}
\end{eqnarray}
Here only two parameters $\chi_1$ and $\chi_3$ appear since the matrix
element has to be symmetric under the exchange of $b$ and $c$ and
the corresponding exchange of initial and final state. The spin
structure of the second order terms of the Lagrangian is the same
and one may write in a similar fashion
\begin{eqnarray}
(-i) \int d^4 x \,
\langle H_b (v) | T \left[ \bar{b}_v \Gamma c_v {\cal K}_c^{(2)} (x) \right]
                   | H_c (v) \rangle &=&
- \Xi_1 \mbox{ Tr } \left\{\bar{M} (v) \Gamma M(v) \right\}
\\ \nonumber
(-i) \int d^4 x \,
\langle H_b (v) | T \left[ \bar{b}_v \Gamma c_v {\cal G}_c^{(2)} (x) \right]
                   | H_c (v) \rangle &=&
- \Xi_3 \mbox{ Tr } \left\{\gamma_\lambda \gamma_5
                     \bar{M} (v) \Gamma s^\lambda M(v) \right\}
\\ \nonumber
(-i)  \int d^4 x \,
\langle H_b (v) | T \left[ \bar{b}_v \Gamma c_v {\cal K}_b^{(2)} (x) \right]
                   | H_c (v) \rangle &=&
- \Xi_1 \mbox{ Tr } \left\{\bar{M} (v) \Gamma M(v) \right\}
\\ \nonumber
(-i)  \int d^4 x \,
\langle H_b (v) | T \left[ \bar{b}_v \Gamma c_v {\cal G}_b^{(2)} (x) \right]
                   | H_c (v) \rangle &=&
- \Xi_3 \mbox{ Tr } \left\{\gamma_\lambda \gamma_5
                     \bar{M} (v) s^\lambda \Gamma M(v) \right\}
\end{eqnarray}
Finally, the double insertions of the first order terms are
parametrized by
\begin{eqnarray}
\frac{(-i)^2}{2} \int d^4 x \,d^4y \,
\langle H_b (v) | T \left[ \bar{b}_v \Gamma c_v {\cal K}_c^{(1)} (x)
                                      {\cal K}_c^{(1)} (y) \right]
                   | H_c (v) \rangle
&=& - A  \mbox{ Tr } \left\{\bar{M} (v) \Gamma M(v) \right\}
\\ \nonumber
(-i)^2 \int d^4 x \,d^4y \,
\langle H_b (v) | T \left[ \bar{b}_v \Gamma c_v {\cal K}_c^{(1)} (x)
                                      {\cal G}_c^{(1)} (y) \right]
                   | H_c (v) \rangle
&=& - B \mbox{ Tr } \left\{\gamma_\lambda \gamma_5
                     \bar{M} (v) \Gamma s^\lambda M(v) \right\}
\\ \nonumber
\frac{(-i)^2}{2} \int d^4 x \,d^4y \,
\langle H_b (v) | T \left[ \bar{b}_v \Gamma c_v {\cal G}_c^{(1)} (x)
                                      {\cal G}_c^{(1)} (y) \right]
                   | H_c (v) \rangle
&=& - \mbox{ Tr } \left\{ C^{\alpha \beta}
\bar{M} (v) \Gamma s_\alpha s_\beta M(v) \right\} \\ \nonumber 
= - C_1 \mbox{ Tr } \left\{
\bar{M} (v) \Gamma M(v) \right\}
  &-& C_3 \mbox{ Tr } \left\{ \gamma_\lambda \gamma_5
\bar{M} (v) \Gamma s^\lambda M(v) \right\} \nonumber
\end{eqnarray}
where we have defined 
\begin{equation}
C_{\alpha \beta} = \frac{1}{3} C_1 (-g_{\alpha \beta } + v_\alpha v_\beta )
  + \frac{i}{2} C_2  \epsilon_{\mu \alpha \beta \lambda} 
                     v^\mu \gamma^\lambda \gamma_5 
\end{equation}

A similar expression is obtained for the double insertion of the first
order Lagrangian of the $b$ quark, involving the same parameters $A$, 
$B$, $C_1$ and $C_3$ due to the exchange symmetry $b \leftrightarrow c$.

Finally, the mixed double insertions need another set of
parameters
\begin{eqnarray}
(-i) \int d^4 x \,d^4y \,
\langle H_b (v) | T \left[ {\cal K}_b^{(1)} (x) \bar{b}_v \Gamma c_v 
                      {\cal K}_c^{(1)} (y) \right]
                   | H_c (v) \rangle
&=& - D  \mbox{ Tr } \left\{\bar{M} (v) \Gamma M(v) \right\}
\\ \nonumber
(-i)^2 \int d^4 x \,d^4y \,
\langle H_b (v) | T \left[ {\cal K}_b^{(1)} (x) \bar{b}_v \Gamma c_v 
                                      {\cal G}_c^{(1)} (y) \right]
                   | H_c (v) \rangle
&=& - E \mbox{ Tr } \left\{\gamma_\lambda \gamma_5
                     \bar{M} (v) \Gamma s^\lambda M(v) \right\}
\\ \nonumber
(-i)^2 \int d^4 x \,d^4y \,
\langle H_b (v) | T \left[ {\cal G}_b^{(1)} (x) \bar{b}_v \Gamma c_v 
                        {\cal K}_c^{(1)} (y) \right]
                   | H_c (v) \rangle
&=& - E \mbox{ Tr } \left\{\gamma_\lambda \gamma_5
                     \bar{M} (v) s^\lambda \Gamma M(v) \right\}
\\ \nonumber
(-i)^2 \int d^4 x \,d^4y \,
\langle H_b (v) | T \left[ {\cal G}_b^{(1)} (x) \bar{b}_v \Gamma c_v 
                        {\cal G}_c^{(1)} (y) \right]
                   | H_c (v) \rangle
&=& - \mbox{ Tr } \left\{ R^{\alpha \beta}
\bar{M} (v) s_\alpha \Gamma s_\beta M(v) \right\}
\end{eqnarray}
where $R$ is given in terms of two parameters
\begin{equation}
R_{\alpha \beta} = \frac{1}{3} R_1 (-g_{\alpha \beta } + v_\alpha v_\beta )
  + \frac{i}{2} R_2  \epsilon_{\mu \alpha \beta \lambda} 
                     v^\mu \gamma^\lambda \gamma_5 
\end{equation}

\subsection{The $0^- \to 0^-$ and $1^- \to 1^-$ Vector Current 
            at Zero Recoil}

In order to obtain the vector current, i.e. the two form factors
$h_+$ and $h_1$, we set $\Gamma = \gamma_\mu$. We shall  
discuss the case of two $0^-$ states keeping the parameter $d_H = 3$ 
explicit. The form factor $h_1$ may then be obtained by setting 
$d_H = -1$ and by replacing the masses $m_B \to m_{B^*}$ and  
$m_D \to m_{D^*}$.  

The matrix element 
of the local term originating from the expansion of the current is 
\begin{equation}
\langle B(v) | \bar{b}_v
\stackrel{\longleftarrow}{(i\slash{D}^\perp)}
\gamma_\mu (i\slash{D}^\perp)  c_v | D(v) \rangle = 
2 v_\mu \sqrt{M_B M_D}  (\lambda_1 + d_H \lambda_2 )
\end{equation}
The traces become 
trivial and one obtains for (\ref{vector})
\begin{eqnarray} \label{vecstat}
&& \langle B(p) | \bar{b} \gamma_\mu c | D(p') \rangle |_{v=v'} = 
2 \sqrt{M_B M_D} \,\, v_\mu 
\left\{ 1 + \left(\frac{1}{2m_c} + \frac{1}{2m_b} \right) 
                       \left( \chi_1 + d_H \chi_3 \right) \right. 
\\ \nonumber
&& +  \left(\left(\frac{1}{2m_c}\right)^2 + \left(\frac{1}{2m_b} \right)^2
      \right) \left( \Xi_1+A+C_1 
                    + d_H [ \Xi_3+B+C_3 ] \right)
\\
&& + \left. \left( \frac{1}{4 m_c m_b} \right) \left(D+R_1- \lambda_1
                                    + d_H [2E+R_2 - \lambda_2 ] \right) 
\right\} \nonumber
\end{eqnarray} 

In order to extract the form factor $h_+$ from this one has to take
into account another trivial source of $1/m$ corrections, which is the
normalization of the states. The right hand side of 
(\ref{vector}) is expressed in terms of the physical meson masses $m_B$ 
and $m_D$. In (\ref{vecstat}) only the masses of the static limit appear
which differ from the physical masses at the order $1/m^2$. To this end, 
one has to take into account a factor 
\begin{equation}
\sqrt{\frac{M_B M_D}{m_B m_D}}
 = 1 + \left(\frac{1}{2m_c^2} +
\frac{1}{2m_b^2} \right) \frac{1}{2} (\lambda_1 + d_H \lambda_2)
\end{equation}
when extracting $h_+$ from (\ref{vecstat}). 

The parameters appearing in (\ref{vecstat})  
are not all independent. The normalization of the flavor diagonal
current is known in full QCD for both matrix elements $0^- \to 0^-$ and 
$1^- \to 1^-$ 
\begin{equation} \label{norm}
\langle B(p) | \bar{b} \gamma_\mu b | B(p) \rangle  = 2 m_B v_\mu =
\langle B^*(p,\epsilon) | \bar{b} \gamma_\mu b | B^*(p,\epsilon) \rangle  
\end{equation}
This may be employed to obtain relations between 
the parameters. Setting $m_b = m_c$, (\ref{norm}) implies the relations  
\begin{eqnarray}
&& \chi_1 = \chi_3 = 0 \label{luke} \\
&& 2 (\Xi_1+A+C_1 + \lambda_1) = -( D+R_1 - \lambda_1) \\
&& 2 (\Xi_3+B+C_3 + \lambda_2) = - (2E+R_2 -\lambda_2) \nonumber
\end{eqnarray}
The first of these equations is Lukes theorem, stating that there
are no first order corrections at the non-recoil point \cite{Lu90}. 
Taking the 
relations between the parameters of the second order 
into account, one obtains for the form factor $h_+$ at 
the non-recoil point 
\begin{equation} 
h_+ (1) = 1 - \left(\frac{1}{2m_c} - \frac{1}{2m_b} \right)^2
              \frac{1}{2} \left(D+R_1 - \lambda_1
                          + 3 [2E+R_2 - \lambda_2 ]\right) 
+ {\cal O} (1/m^3)
\end{equation}
Similarly, by the appropriate replacements one obtains
\begin{equation} 
h_1 (1) = 1 - \left(\frac{1}{2m_c} - \frac{1}{2m_b} \right)^2
              \frac{1}{2} \left(D+R_1 - \lambda_1
                          - [2E+R_2 - \lambda_2 ]\right) 
+ {\cal O} (1/m^3)
\end{equation}

Looking at the definition of the parameters entering the $1/m^2$
corrections it turns out that to order $1/m^2$ the only 
input need are the two parameters $\lambda_1$ and $\lambda_2$ from 
the local dimension 5 operators and the time ordered product involving
insertions of both ${\cal L}_b^{(1)}$ and ${\cal L}_c^{(1)}$, which is 
given in terms of four parameters. The results for $h_+$ and $h_1$ may 
also be written as 
\begin{eqnarray} \label{hplus}
h_+ (1) &=&  1  - \left(\frac{1}{2m_c} - \frac{1}{2m_b} \right)^2
   \frac{1}{2} \left( -\lambda_1 - 3 \lambda_2  \vphantom{\int} \right.
\\ \nonumber  &+&  \left. 
 (-i)^2 \frac{1}{2\sqrt{M_B M_D}} \int d^4 x \,d^4y \,
\langle B(v) | T \left[ {\cal L}_b^{(1)} (x) \bar{b}_v c_v 
                                      {\cal L}_c^{(1)} (y) \right]
                   | D(v) \rangle \right)
+ {\cal O} (1/m^3) \\ 
h_1 (1) &=&  1  - \left(\frac{1}{2m_c} - \frac{1}{2m_b} \right)^2
   \frac{1}{2} \left( -\lambda_1 + \lambda_2  \vphantom{\int} \right.
\\ \nonumber  &+&   \left. 
(-i)^2 \frac{1}{2\sqrt{M_B M_D}} \int d^4 x \,d^4y \,
\langle B^*(v,\epsilon) | T \left[ {\cal L}_b^{(1)} (x) \bar{b}_v c_v 
                                      {\cal L}_c^{(1)} (y) \right]
                   | D^*(v,\epsilon) \rangle \right)
+ {\cal O} (1/m^3)  
\end{eqnarray}
This relation has a simple intuitive interpretation. The contributions from 
the local dimension five operators ($\lambda_1$ and $\lambda_2$)
originate from the matching of the 
field operators of full QCD to the ones of the effective theory to order 
$1/m_c$  and $1/m_b$ respectively. However, also the states receive 
corrections and the matrix element involving the time ordered 
product corresponds to 
the corrections to the states to order $1/m_b$ and $1/m_c$ respectively.
The local and the nonlocal contributions are of order $1/(4m_c m_b)$,  
but due to the normalization of the 
matrix element for the flavor diagonal case all other terms of order 
$1/m_b^2$ and $1/m_c^2$ have to be related to these such that  
the normalization is preserved in the case $m_c = m_b$. This fact 
leads to the prefactor $(1/m_c - 1/m_b)^2$ in front of the correction term
of $h_+ (1)$. 

This result agrees with the one found in \cite{FN92}. In particular, 
one may see that at the non-recoil point the form factors may be 
expressed in terms of the parameters $D_1$, $D_3$, $D_4$, $D_5$ and 
$D_6$ defined in  \cite{FN92}. However, the 
representation in terms of the Pauli matrices reveals a relation 
between the parameters of \cite{FN92}, at least at the 
non-recoil point. In total there are six 
independent parameters $\lambda_1$, $\lambda_2$, $D$, $E$ $R_1$ 
and $R_2$ at $v=v'$ and one may show that $R_1 = 3 (D_4 + D_5)$ and 
$R_2 = 2 (D_5 + D_6)$. 

The $1/m_Q^2$ corrections to $h_+$ and $h_1$ depend on the spin 
symmetry conserving contribution $X = D + R_1 - \lambda_1$ and on 
the spin symmetry breaking combination $Y = 2E + R_2 - \lambda_2$.
If one considers in addition the form factor $h_{A1}$ then a third 
combination of the six parameters $\lambda_1$, $\lambda_2$, $D$, 
$E$ $R_1$ and $R_2$ is needed.   

In order to perform a model independent extraction of $V_{cb}$ from 
the decay $B \to D \ell \nu$ one has to take into account also the 
second form factor $h_-$ of the vector current. However, considering
only forward matrix elements one cannot say anything about this form 
factor, and one has to consider different velocities along the lines 
of \cite{FN92}.

\subsection{The $0^- \to 1^-$ Axial Vector Current at Zero Recoil}

For a model independent extraction of $V_{cb}$  
the process $B \to D^* e \nu$ is much more interesting than
$B \to D \ell \nu$. 
The relevant form factor is $h_{A1}$ as defined in (\ref{axial}). 
To leading order we have at the non-recoil point due to 
spin symmetry
\begin{equation}
\langle B(v) | \bar{b}_v  c_v | D (v) \rangle  = 2 \sqrt{M_B M_D} = 
- \langle B(v) | \bar{b}_v (s \epsilon) c_v | D^* (v, \epsilon) \rangle   ,
\end{equation}
and thus $h_{A1}$ is normalized in the same way as $h_+$ and $h_1$.
To subleading order $h_{A1}$ will differ from both $h_+$ and $h_1$ due to 
spin symmetry breaking. This is, however, calculable in terms of
the parameters which have been introduced above. 

The local dimension five terms may be expressed in terms of 
$\lambda_1$ and $\lambda_2$
\begin{equation}
\langle B(v) | \bar{b}_v 
\stackrel{\longleftarrow}{(i\slash{D}^\perp)}
\gamma_\mu \gamma_5 (i\slash{D}^\perp)  c_v | D^*(v,\epsilon) \rangle = 
2 \sqrt{M_B M_D} \, \epsilon_\mu \left(- \frac{1}{3}\lambda_1 
                 -  \lambda_2\right) 
\end{equation}
while the contributions from the $T$-products are evaluated by 
replacing $\Gamma \to \gamma_\mu \gamma_5$ and using the representation 
matrix for the vector meson in the final state. Taking into account the 
contributions form the normalization of the states one obtains for 
the form factor $h_{A1}$ at zero recoil
\begin{eqnarray}
h_{A1} (1) &=& 1 - \left(\frac{1}{2m_b} \right)^2
               \frac{1}{2} \left(D+R_1 - \lambda_1
                          + 3 [2E+R_2 - \lambda_2 ]\right)
\\ \nonumber 
               &-& \left(\frac{1}{2m_c} \right)^2
              \frac{1}{2} \left(D+R_1 - \lambda_1
                          - [2E+R_2 - \lambda_2 ]\right) 
\\ \nonumber 
               &+& \left(\frac{1}{4m_b m_c} \right) 
                 \left(D + 2E - \frac{1}{3} R_1 - R_2 
                              + \frac{1}{3} \lambda_1 + \lambda_2 \right)
+ {\cal O} (1/m^3)
\end{eqnarray}
The structure of this result may be understood from spin symmetry. 
In the heavy quark limit, spin symmetry relates all the form factors
$h_+$, $h_1$ and $h_{A1}$. If one would take into account only the 
corrections of order $1/m_b^n$, then the spin symmetry of the $c$ 
quark would still be unbroken and one may rotate the $D^*$ meson 
into a $D$ meson. Hence the $1/m_b^2$ corrections to $h_{A1}$ have 
to be the same as the $1/m_b^2$ corrections to $h_+$. Similarly, 
and more importantly, the $1/m_c^2$ corrections to $h_{A1}$ have 
to be the same as the $1/m_c^2$ corrections to $h_1$ since we may 
now use the spin symmetry of the $b$ quark to rotate the $B$ meson 
into a $B^*$ meson. Finally, the mixed insertions break both spin 
symmetries and thus cannot be expressed in terms of $h_1$ or 
$h_+$. 

In total, the three form factors may be reexpressed in terms 
of three parameters $X$, $Y$ and $Z$ 
\begin{eqnarray}
h_+ &=& 1 - \left( \frac{1}{m_b} - \frac{1}{m_c} \right)^2
            \frac{1}{2} (X + 3 Y)  \\
h_1 &=& 1 - \left( \frac{1}{m_b} - \frac{1}{m_c} \right)^2
            \frac{1}{2} (X - Y)  \\
h_{A1} &=&  1 - \left( \frac{1}{m_b} \right)^2
                 \frac{1}{2} (X + 3 Y) -
                 \left( \frac{1}{m_c} \right)^2
                 \frac{1}{2} (X - Y)  
             +   \frac{1}{4 m_b m_c}  
                 \left( -\frac{1}{3} X - Y +Z \right) 
\end{eqnarray}
where
\begin{equation}
X =  D + R_1 - \lambda_1, \quad 
Y =  2E + R_3 - \lambda_2, \mbox{ and } 
Z = \frac{4}{3} D + 4 E 
\end{equation} 
where $X$ corresponds to the spin symmetry conserving interactions, 
$Y$ to the  spin symmetry breaking ones, while $Z$ is a mixture of 
spin symmetry conserving and spin symmetry breaking terms.

\subsection{Discussion of the Results and Quantitative Estimates}
Finally we shall discuss the results and try to give a numerical
estimate of the corrections. In general, this needs input beyond 
heavy quark effective theory as e.g. a model. A few things, however, 
may be said with two plausible assumptions.

The form factors $h_+$ and $h_1$ are form factors which are in the 
heavy quark limit related to matrix elements of conserved currents,
which generate heavy flavor symmetry. The operators 
\begin{equation}
K_+ = \int d^3  \vec{x} \,\, \bar{b}_v c_v ,\quad
K_- = \int d^3 \vec{x} \,\, \bar{c}_v b_v ,\quad
K_0 = \int d^3 \vec{x} \,\, [ \bar{b}_v b_v  - \bar{c}_v c_v ]
\end{equation}
are generators of the heavy flavor symmetry satisfying 
$[K_+ \, , \, K_-] = K_0$, and hence one derives in the symmetry 
limit $ \langle B(v) | Q_+ | D (v) \rangle = \sqrt{M_B M_D}$ and 
a similar relation for two vector mesons, implying that $h_+ (1) = 
h_1 (1)= 1$.  

In the presence of explicit symmetry breaking one splits the Hamiltonian
in a symmetry conserving piece $H_0$ and a symmetry breaking term 
$\lambda H_{SB}$, such that $[K_j \, , \, H_0] = 0$. Since now 
$[K_j \, , \, H] = \lambda [K_j \, , \, H_{SB}] \neq 0$ the generators 
$K_j$ become time dependent and one has at $x_0 = 0$
\begin{equation} \label{AGT}
1 = \left| \frac{\langle B(v) | Q_+ | D (v) \rangle}
                {2 \sqrt{M_B M_D}} \right|^2 
  + \sum_{X \neq D} \left( \frac{\lambda}{E_B - E_X} \right) ^2
    \left| \frac{\langle B(v) | Q_+ | X (v) \rangle }
                {2 \sqrt{M_B M_X}} \right|^2
\end{equation}
where $X (v)$ is a hadronic state in which the $c$ quark moves with
velocity $v$. On the left hand side we have neglected the matrix element
of $\bar{c}_v c_v$ between the $B$ meson states. 

Relation (\ref{AGT}) is the standard derivation of the Ademollo Gatto 
theorem \cite{AGTh,FL65}, and it allows two observations. The state $X$ is 
not in the lowest spin symmetry doublet and hence the 
energy difference $E_B - E_X$ is not  
of the order $\lambda$, but of the order 1. Hence the corrections due
to symmetry breaking are of second order in the symmetry breaking 
interaction, which is the well known Ademollo Gatto theorem \cite{AGTh}. 

Secondly, and more importantly for the present discussion, it shows 
that one expects that 
$$
\left| \frac{\langle B(v) | Q_+ | D (v) \rangle}{2 M_B M_D} \right|^2 \le 1 ,
$$ 
since the sum on the right hand side of (\ref{AGT}) is positive. 
This means that $h_+ - 1 \le  0$ and $h_1 - 1 \le 0$. 

However, it is known that short distance contributions may change  
this conclusion. For instance, the full one loop QCD calculation 
yields for $h_+ (1)$ \cite{Ne92}
\begin{equation} \label{sdc}
h_+ (1) = 1 + \frac{\alpha}{\pi} \left( \frac{m_b + m_c}{m_b - m_c} 
              \ln \left(\frac{m_b}{m_c}\right) - 2 \right)
\end{equation}
which yields a positive contribution to the normalization. This 
may be traced back to the matrix element of $\bar{c}_v c_v$
which we have neglected on the left hand side of (\ref{AGT}). 
We shall assume in the following that the positive short distance
contribution (\ref{sdc}) is compensated by the long distance one, for 
which (\ref{AGT}) holds.

\begin{figure}[t]
   \vspace{0.5cm}
   \epsfysize=9cm
   \centerline{\epsffile{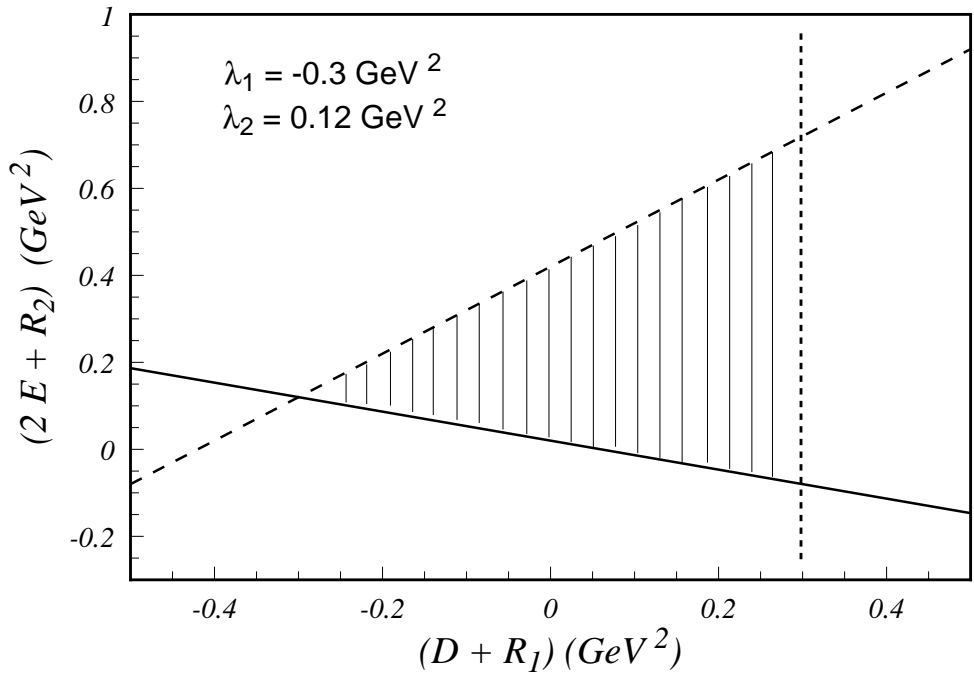}}
   \centerline{\parbox{11cm}{\caption{\label{fig}
Allowed region for the spin symmetry conserving $(D + R_1)$ and 
spin symmetry breaking terms $(2E + R_2)$ of the time ordered product.
The solid (dashed) line is from the constraint $h_+ (1) < 1$ 
($h_1 (1) < 1$), while the vertical dotted line is the assumed upper limit 
for $(D + R_1)$.
  }}}
\end{figure}

From this requirement one obtains two constraints for the parameters
\begin{eqnarray} \label{tpcon}
(D + R_1) - \lambda_1 + 3 (2E + R_2) - 3 \lambda_2 &>& 0   \\
(D + R_1) - \lambda_1 - (2E + R_2) + \lambda_2     &>& 0  ,  
\end{eqnarray}
which are equivalent to $(D + R_1) \ge \lambda_1$ 
and $ - (D + R_1  - \lambda_1) / 3 \le  (2E + R_2 \lambda_2) 
\le (D + R_1  - \lambda_1) $. 

In fig.\ref{fig} we plot the spin symmetry conserving contribution
of the time ordered products $(D + R_1)$ 
versus the spin symmetry breaking part $(2E + R_2)$. The allowed region 
is the one below the dashed and above the solid line, where 
$h_1 - 1 < 0$ and $h_+ - 1 < 0$. 

In order to obtain some numerical estimate we shall assume that 
$(D + R_1) \le - \lambda_1$ which should be a reasonable order of 
magnitude for the spin symmetry conserving terms of the time ordered
products. Thus the parameters for the time ordered product terms should 
lie within the shaded triangular region in fig.\ref{fig}. 

We shall estimate the contributions to $h_{A1} (1)$ by observing that, 
numerically, the contributions of the order $1/m_c^2$ are by far the 
largest. As argued above, spin symmetry enforces that 
these contributions are the same as the
ones to the form factor $h_1$. Maximizing the 
form factor $h_+$ in the shaded triangle of fig.\ref{fig}, we
have for $h_{A1} (1)$ 
\begin{equation}
- \left( \frac{1}{2 m_c} \right)^2 
  \left(-\frac{4}{3} \lambda_1 \right) \le h_{A1} (1) - 1 \le 0
\end{equation}
Using $\lambda_1 \sim - 0.3$, corresponding to 
the minimal value assumed here, 
one obtains corrections to $h_{A1} - 1$ ranging between zero and
-5\%. This is consistent with the estimate performed in \cite{FN92}
based on a simple wave function overlap model, once updated values 
for the parameters are used \cite{Nover}.

The present estimate is not based on a model but on the assumption 
that the spin symmetry conserving contributions to the time ordered 
products satisfy $\lambda_1 \le (D + R_1) \le - \alpha \lambda_1$ 
with $\alpha \sim {\cal O} (1)$. The dependence on $\alpha$ is not 
very strong; for $\alpha = 2$ one obtains $h_{A1} - 1 > - 6\%$ and 
$1/m_Q^2$ corrections to $h_{A1}$ exceeding 8\% are very 
unlikely.

\section{Conclusions}
In this paper we have considered forward matrix elements
of local operators of higher dimension and their time ordered products 
with terms originating form the heavy mass expansion of the Lagrangian.
Due to the projection $P_+ = (\slash{v}+1)/2$ appearing in heavy 
quark effective theory the Dirac algebra simplifies and only two 
types of matrix elements of local operators appear. In 
addition, the spin structure of the time ordered products of these 
operators with higher order terms form the Lagrangian may be analyzed in 
a simple way.  
 
Matrix elements of this type appear in two important applications. 
Performing a heavy mass expansion for inclusive decays along the lines
of Bigi et al. \cite{Bigi} these matrix elements parametrize the 
non-perturbative input required beyond the leading order in the $1/m_Q$
expansion of total rates as well as for inclusive decay spectra.   

The second application are the form factors for weak transitions at 
the non-recoil point. The symmetries of the heavy quark limit yield 
the normalization of the weak transitions between heavy quarks; this 
fact may be employed to perform a model independent determination of 
$|V_{cb}|$. The recoil corrections to the normalization are given in 
terms of the forward matrix elements considered here. At the non-recoil 
point the analysis simplifies drastically, mainly due to the 
simplification of the Dirac algebra, as compared to the general analysis.  
 
As an example we have reconsidered the second order corrections to the 
semileptonic transition $B \to D^{(*)}$. These corrections have been 
studied already in \cite{FN92} for the general case. At the non-recoil 
point the present analysis agrees with the one performed in \cite{FN92}.
However, it turns out that at $v=v'$ some of the parameters given in 
\cite{FN92} are in fact not independent. 

The second order corrections of to the weak decay form factors are
all parametrized in terms of five matrix elements
$$ 
\lambda_1 = \frac{1}{2M_Q}
\langle H(v) |\bar Q_v (iD)^2 Q_v| H(v) \rangle ,  
\qquad
\lambda_2 = \frac{1}{2M_Q}
\langle H(v) |\bar Q_v (iD_\alpha) (iD_\beta) 
(-i\sigma^{\alpha \beta}) Q_v| H(v) \rangle 
$$
and the matrix elements of double insertions of the first order 
correction to Lagrangian 
\begin{eqnarray*}
&& (-i) \frac{1}{2\sqrt{M_B M_D}} \int d^4 x \,d^4y \,
\langle B(v) | T \left[ {\cal K}_b^{(1)} (x) \bar{b}_v c_v 
                                      {\cal K}_c^{(1)} (y) \right]
                   | D(v) \rangle  \\
&& (-i) \frac{1}{2\sqrt{M_B M_D}} \int d^4 x \,d^4y \,
\langle B(v) | T \left[ {\cal G}_b^{(1)} (x) \bar{b}_v c_v 
                                      {\cal K}_c^{(1)} (y) \right]
                   | D(v) \rangle \\
&& (-i) \frac{1}{2\sqrt{M_B M_D}} \int d^4 x \,d^4y \,
\langle B(v) | T \left[ {\cal G}_b^{(1)} (x) \bar{b}_v c_v 
                                      {\cal G}_c^{(1)} (y) \right]
                   | D(v) \rangle 
\end{eqnarray*}
All other matrix elements of time ordered products are related 
to these by the normalization condition for the vector current 
in the full theory.
 
All these matrix elements are non-perturbative. 
In principle 
they may be measured on the lattice and first results have been 
reported \cite{Lattice}. However, in the meantime one has to rely 
on some model 
to estimate their size. In the present paper we have used a reasonable
guess for the spin symmetry conserving contributions of the time 
ordered products to get some upper limit for the $1/m_Q$ corrections
to $h_{A1}$ at the non-recoil point. The main result of this analysis
is that the corrections to $h_{A1} (1)$ are small, ranging between 
$-5\%$ and zero. 

Including also the leading and subleading  QCD radiative 
corrections \cite{corr} to the 
normalization of $h_{A1}$ one concludes that 
\begin{equation}
h_{A1} (1) = 0.96 \pm 0.03 , 
\end{equation}
and thus the corrections at zero recoil are small.

\section*{Acknowledgements}
It is a pleasure to thank M. Neubert for help in the comparison 
with \cite{FN92}, and A. Ali and N. Uraltsev
for stimulating discussions.


\begin{thebibliography}{99}
\bibitem{IW89}{N. Isgur and M. Wise,
               Phys. Lett. {\bf B232} (1989) 113;
               Phys. Lett. {\bf B237} (1990) 527.}
\bibitem{PW88}{H. Politzer and M. Wise,
               Phys. Lett. {\bf B206} (1988) 681;
               Phys. Lett. {\bf B208} (1988) 504.}
\bibitem{VS87}{M. Voloshin and M. Shifman,
               Sov. J. Nucl. Phys. {\bf 45} (1987) 292;
               Sov. J. Nucl. Phys. {\bf 47} (1988) 511.}
\bibitem{EH90}{E. Eichten and B. Hill,
               Phys. Lett. {\bf B234} (1990) 511.}
\bibitem{Ne93}{A detailed review is given in:
               M. Neubert,
               SLAC-PUB-6263 (to appear in Phys. Rep.).}
\bibitem{Gr90}{B. Grinstein,
               Nucl. Phys. {\bf B339} (1990) 253.}
\bibitem{Ge90}{H. Georgi,
               Phys. Lett. {\bf B240} (1990) 447.}
\bibitem{FG90}{A. Falk, H. Georgi, B. Grinstein and M. Wise,
               Nucl. Phys. {\bf B343} (1990) 1.}
\bibitem{MRR}{T. Mannel, W. Roberts and Z. Ryzak,
               Nucl.\ Phys.\ {\bf B 368} (1992) 204.}
\bibitem{corr}{A detailed list of references on the QCD radiative 
               corrections as well as on the recoil corrections may 
               be found in \cite{Ne93}.}
\bibitem{Lu90}{M. Luke, 
               Phys. Lett. {\bf B252} (1990) 447.}
\bibitem{FN92}{A. Falk and M. Neubert, 
               Phys.\ Rev.\ {\bf D 47} (1993) 2965.}
\bibitem{CGG90}{J. Chay, H, Georgi and B. Grinstein,
               Phys. Lett. {\bf B247} (1990) 399. }
\bibitem{Bigi}{I. Bigi {\it et al.}, 
               Phys.\ Rev.\ Lett.\ {\bf 71} (1993) 496.}
\bibitem{Blok}{B. Blok {\it et al.},
                Preprint TPI-MINN-93/33-T (1993).}
\bibitem{WM93}{M. Wise and A Manohar, 
                Preprint UCSD/PTH 93-14 (1993).} 
\bibitem{Ma93}{T. Mannel, Preprint IKDA 93/26, to appear in 
               Nucl.\ Phys.\ {\bf B}.}
\bibitem{FGL92}{A. Falk, B. Grinstein and M. Luke, 
               Nucl. Phys. {\bf B 357} (1991) 185.}
\bibitem{Br92}{P. Ball and V. Braun, Preprint MPI-Ph/93-51 (1993).}
\bibitem{Npriv}{M. Neubert, private communication.}
\bibitem{Bmotion}{I. Bigi {\it et al.}, Preprint CERN-TH.7129/93 (1993).}
\bibitem{FLS93}{A. Falk, M. Luke and M. Savage, 
                SLAC preprint SLAC-PUB-6317 (1993).}
\bibitem{AGTh}{M. Ademollo and R. Gatto, 
               Phys.\ Rev.\ Lett.\ {\bf 13} (1964) 262.}
\bibitem{FL65}{G. Furlan {\it et al.},
               Nuovo Cim. {\bf 38} (1965) 1747.}
\bibitem{Ne92}{M. Neubert, 
               Nucl. Phys. {\bf B 371} (1992) 149.}
\bibitem{Nover}{The updated numbers will appear in \cite{Ne93}.} 
\bibitem{Lattice}{P. MacKenzie, Review talk at the 16th 
                  International Symposium on Lepton and 
                  Photon Interactions,  
                  Ithaca, NY, 10-15 Aug 1993, 
                  FERMILAB-CONF-93-343-T, (1993).}      
\end{thebibliography}
\end{document}